\documentclass{natureprintstyle}
\usepackage{amssymb,amsmath,graphicx}
\def\ltsima{$\; \buildrel < \over \sim \;$}
\def\lsim{\lower.5ex\hbox{\ltsima}}
\def\gtsima{$\; \buildrel > \over \sim \;$}
\def\gsim{\lower.5ex\hbox{\gtsima}}

\newcommand{\be}{\begin{equation}}
\newcommand{\en}{\end{equation}}

\def\cmdue {\rm \ cm^{-2}}

\def\msole {~M_{\odot}}

\begin{document}

\title{A minor body falling onto a neutron star as an explanation for the unusual $\gamma$-ray burst GRB 101225A
}

\author{S. Campana$^1$,
G. Lodato$^2$, P. D'Avanzo$^1$, N. Panagia$^{3,4,5}$, E. M. Rossi$^6$, M. Della Valle$^7$, G. Tagliaferri$^1$, 
L. A. Antonelli$^8$, S. Covino$^1$, G. Ghirlanda$^1$, G. Ghisellini$^1$, A. Melandri$^1$, E. Pian$^{9,10}$, 
R. Salvaterra$^{11}$,  G. Cusumano$^{12}$, V. D'Elia$^{13,8}$, D. Fugazza$^1$, E. Palazzi$^{14}$, B. Sbarufatti$^1$, 
S. D. Vergani$^1$ 
}

\maketitle

\begin{affiliations}
\item INAF - Osservatorio astronomico di Brera, Via E. Bianchi 46,  I-23807
Merate (LC), Italy.

\item Dipartimento di Fisica, Universit\`a degli Studi di Milano, Via Celoria 16, I-20133 Milano, Italy.

\item  Space Telescope Science Institute, 3700 San Martin Drive, Baltimore, MD 21218, USA.

\item  INAF - Osservatorio astrofisico di Catania, Via Santa Sofia 78, I-95123 Catania, Italy.

\item Supernova Ltd, OYV \#131, Northsound Road, Virgin Gorda, British Virgin Islands.

\item  Leiden Observatory, Leiden University, P.O. Box 9513, 2300 RA, Leiden, The
Netherlands.

\item INAF - Osservatorio astronomico di Capodimonte, Salita Moiariello 16, I-80131 Napoli, Italy.

\item  INAF - Osservatorio astronomico di Roma, Via Frascati 33, I-00044 Monte Porzio (Roma), Italy.

\item  Scuola Normale Superiore, Piazza dei Cavalieri 7, I-56126 Pisa, Italy.

\item  INAF - Osservatorio astronomico di Trieste, Via G. Tiepolo 11, I-34143 Trieste, Italy.

\item  Dipartimento di Fisica e Matematica, Universit\`a dell'Insubria, Via Valleggio 7, I-22100 Como, Italy.

\item  INAF - Istituto di Astrofisica Spaziale e Fisica Cosmica, Via U. La Malfa 153, I-90146 Palermo, Italy.

\item  ASI Science Data Centre, Via Galileo Galilei, I-00044 Frascati (Roma), Italy.

\item INAF - Istituto di Astrofisica Spaziale e Fisica Cosmica, Sezione di Bologna, Via Gobetti 101, I-40129 Bologna, Italy.
\end{affiliations}

\begin{abstract} 
The tidal disruption of a solar mass star around a supermassive black hole has been extensively studied analytically$^{1,2}$
and numerically$^3$.
In these events the star develops into an elongated banana-shaped structure.
After completing an eccentric orbit, the bound debris fall onto the black hole, 
forming an accretion disk and emitting radiation$^{4,5,6}$.
The same process may occur on planetary scales, if a minor body orbits too close to its star. 
In the Solar System, comets fall directly onto our Sun$^7$
or onto planets$^8$.
If the star is a compact object, the minor body can become 
tidally disrupted. Indeed, one of the first mechanisms invoked to produce
strong $\gamma-$ray emission involved accretion of comets onto neutron
stars in our Galaxy$^9$.
Here we report that the peculiarities of the 
`Christmas' $\gamma-$ray burst  (GRB 101225A$^{10}$)
can be explained by a tidal disruption event of a minor body around a Galactic isolated neutron star.
This result would indicate either that minor bodies can be captured by compact stellar remnants more 
frequently than it occurs in our Solar System or that minor body formation is relatively easy around 
millisecond radio pulsars.
A peculiar Supernova associated to a GRB may provide an alternative explanation$^{11}$.
\end{abstract}

GRB 101225A image-triggered the Burst Alert Telescope (BAT) onboard the NASA's Swift satellite
on 25.776 December 2010 UT.
The event was extremely long, with a $T_{90}$ (the time interval during which $90\%$ of the flux was emitted) 
$> 1.7$ ks, and smooth$^{10}$ 
by comparison with typical $\gamma-$ray bursts$^{12}$ (GRBs).
The total 15--150 keV fluence recorded by the BAT 
is $\gsim 3\times 10^{-6}$ erg cm$^{-2}$  and there are no signs of decay.
The X-ray Telescope and the UltraViolet Optical Telescope on board Swift 
found a bright, long-lasting counterpart to the GRB. 
Strong variability is observed in the early X--ray light curve.
The optical counterpart, which was detected in all of the UltraViolet Optical Telescope filters, 
lags the X--ray light curve (Fig. 1).
The X--ray and optical light curves are reminiscent a shock break-out event observed 
in association with GRB060218$^{13}$,
but are fainter ($\sim 3.5$ mag, i.e. a factor of $\sim 25$) and do not have an X--ray afterglow at 
later times or a bright supernova component.
Measurements by the European Space Agency's XMM-Newton space observatory 
failed to detect an afterglow with an upper flux limit of $\sim 10^{-14}$ erg cm$^{-2}$ s$^{-1}$ 
at the $3\,\sigma$ confidence level (0.5--10 keV, observation made at 
$\Delta t=23$ d after the trigger).

Ground-based telescopes also followed the event, mainly in the $R$ and $I$ bands 
(Fig. 1). Surprisingly, optical spectra recorded in the first few days after the event 
failed to detect any spectral feature, instead revealing a smooth blue continuum$^{14,11}$. 
At later times, the optical light curves revealed a color change
from blue to red. The NASA Hubble Space Telescope imaged the field at $\Delta t=20$ d, finding a quite 
red object ($V_{606W}(AB)=24.6$ and $V_{435W}-B_{606W}=1.6$), with no host galaxy  apparent at 
magnitudes at most 1 mag fainter$^{15}$. 
Later observations carried out at the INAF Telescopio Nazionale Galileo 
recorded an $I-$band magnitude of $I=23.64\pm0.24$ at $\Delta t=44$ d but recorded a non-detection 
($I>24.5$, $3\,\sigma$ confidence level) at  $\Delta t=55$ d. Observations 
at the Gran Telescope Canarias ($\Delta t=180$ d) detected GRB 101225A at 
$g'_{\rm AB}=27.21\pm0.27$ and $r'_{\rm AB}=26.90\pm0.14$ (ref. 11). 

These observational characteristics make GRB 101225A unique. Without a-priori knowledge of the distance, this 
event can result from different progenitors. We explored several scenarios (Supplementary Information).
Extragalactic models (involving GRBs, supernovae or blazars) are unlikely on the basis of total energy arguments, 
in combination with the faintness of the optical counterpart and putative host galaxy. 
Galactic models (involving X--ray binaries, magnetars or
flare stars) are ruled out as well, on the basis of the extremely faint quiescent counterpart and on the relatively high Galactic latitude.
The timescale of the event, the longer 
duration at longer wavelengths and the faintness of a quiescent counterpart at all wavelengths all suggest a 
disruptive event involving a compact object. 
We propose that GRB 101225A results from the tidal disruption of a minor body around an isolated neutron star.
Similar models have been already put forward to explain the presence of metal absorption lines in the white 
dwarf optical spectra$^{16}$ 
and the infrared excess in the spectra of hot white dwarfs$^{17}$.

A minor body  gets disrupted if it comes within a distance of $\sim 10^5-10^6$ km  from the 
neutron star, where the internal forces that hold the body together 
are overwhelmed by the tidal pull of the neutron star. The debris that remain bound 
are thrown into highly eccentric orbits and then fall back to form an accretion disk around the star$^{18,19}$.
If the minor body has a low tensile strength it gets fully disrupted, 
its early behaviour can be chaotic owing to the formation of the disk with 
flares coincident with the periastron passages$^{6,20}$.
and the bolometric light curve is expected to decay, in the long term, as $t^{-5/3}$ (ref. 1--2),

If the matter that falls back to form the disk accretes at a sub-Eddington rate at all times (as in our case), 
the disk emission can be described as a spatial superposition of black bodies 
of increasing temperatures and depends on the parameters of the encounter$^{21}$.
The physics of the event depends, to a first approximation, only on a very few parameters: the periastron
of the minor body's orbit ($r_{\rm p}$),
the minor body's mass and radius ($M_*$ and $R_*$; the mean density of a minor body in the Solar System is in the range 
$\sim 1-10$ g cm$^{-3}$, which adds an additional constraint), the minor body's tensile strength, 
the neutron star mass ($M_{\rm NS}$) and the source distance.

We fit the X-ray (1 keV), ultraviolet ($UVW2$ and $UVW1$ bands, centred at 2,030 and 2,634\,\AA, respectively) and optical 
($R$ and $I$ bands, centred at 6,400 and 7,700\,\AA, respectively) light curves of  GRB 101225A with a four-parameters phenomenological 
model of tidal disruption onto a neutron star$^{21}$ (Fig. 1).
We assume that the neutron star mass is $1.4\msole$ and that the minor body has a density 
of 1 g cm$^{-3}$ and a low internal tensile strength. 
Our model is indicative of the physics involved, even though it cannot capture the full details of the event.
We find an adequate fit for $M_*\sim 5\times 10^{20}$ g  (half the mass of the asteroid Ceres)
and a periastron radius of $\sim 9,000$ km (which is well within the tidal radius, thus demonstrating the 
internal consistency of the model). The mass of the minor body can be estimated within a factor of $\lsim 3$.
This mass is large by comparison with those of objects in the asteroid belt but is typical of objects in the Kuiper belt.
 
Assuming that half of the mass is accreted onto the neutron star and the total fluence is $4\times 10^{-5}$ erg 
cm$^{-2}$, estimated within the $1-10^5$ s time interval (including a bolometric correction motivated by spectral fitting),
we derive a distance of $\sim 3$ kpc, placing GRB 101225A in the Perseus arm of the Galaxy.
The observed radiation is emitted both by an accretion disk (with an outer radius twice as large as periastron) 
and during the impact of matter onto the neutron star surface.
The early spectral energy distribution (SED) at five different epochs can be adequately fitted by the emission from 
an accretion disk (with a temperature of $T_{\rm d}\sim 1.8$ keV in units of energy, in agreement with model predictions) plus the 
emission from a boundary layer in which the disk matter slows down to accrete onto the neutron star
(Supplementary Information). In particular, the initial high-energy emission observed by the BAT can be modelled as a 
black body with $T\sim 10$ keV$^{11}$. This cannot come from the disk but can be easily accounted for
by mass accretion onto the neutron star surface. 
The presence of the boundary layer component rules out the compact object's being a black hole.
Small differences in the return times of debris are expected during the first periastron passages, and these give rise 
to X--ray variability on the timescale of this fall back$^{6,20,22}$, 
as observed in the X--ray band (Fig. 1, inset).
The X--ray variability timescale is indeed very similar to the fall-back timescale obtained
from our modeling.

The phenomenological model can naturally explain the (observed) slower decay at longer wavelengths. 
It somewhat underestimates the observed decay of the optical flux starting at $\sim 20$ d.
Changes in the properties of the accretion disk are expected on the basis of the theory of accretion onto neutron stars.
The inner disk radius is increased by the neutron star's magnetic field (even when this takes its minimal value, $10^8-10^9$ G).
Furthermore, when the accretion rate decreases below a critical rate, the disk develops a well-known thermal-viscous 
instability leading to a fast cooling$^{23,24}$. 
We have calculated that, on the basis of the predicted evolution of the mass inflow rate, this transition occurs at $\Delta t\sim 15-21$  d.
The $\Delta t=40$ d SED can be modelled as arising from a cool ring ($T\sim 4500$ K) with a large inner radius. 
The $\Delta t=180$ d SED can be interpreted as the emission coming from a larger and colder ring ($T\sim 650$ K, 
resembling a protoplanetary disk) plus a contribution from the neutron star heated by 
the accreted material (Supplementary Information).

Estimating the rate of such events is challenging. A wandering neutron star can intercept minor 
bodies as it passes through a planetary system. 
The capture rate can be evaluated on the basis of the hypothesis that all stars have Oort clouds similar to that of 
the Sun$^{25}$. Applying this to the case of GRB 101225A, 
we derive a rate of $\sim 0.3$ events per year, in line with Swift observations.
However, indirect arguments suggest a lower capture rate$^{26}$ and, in addition, larger minor bodies 
are rarer.
The likelihood of a neutron star retaining its original population of small bodies is rather low, because they are 
unlikely to survive the supernova event. Nevertheless, planetary-like systems can reform around millisecond radio 
pulsars$^{27,28}$, providing a viable mechanism for the occurrence of GRB 101225A.

\bigskip

{\bf References}

\noindent \hangindent2em  \hangafter=1  1.\ Rees, M. J.,
Tidal disruption of stars by black holes of 10 to the 6th-10 to the 8th solar masses in nearby galaxies.
Nature, {\bf 333}, 523--528 (1988).

\noindent  \hangindent2em  \hangafter=1  2.\ Phinney, E. S., 
Manifestations of a Massive Black Hole in the Galactic Center.
proc. of the 136th IAU Symp. `The Center of the Galaxy', M. Morris ed., Kluwer, Dordrecht, p.543--553 (1989).

\noindent  \hangindent2em  \hangafter=1  3.\ Evans, C. R., Kochanek, C. S.,	
The tidal disruption of a star by a massive black hole. 
Astrophysical Journal, {\bf 346}, L13--L16 (1989).

\noindent \hangindent2em  \hangafter=1   4.\ Renzini, A., Greggio, L., di Serego Alighieri, S., Cappellari, M., Burstein, D., Bertola, F.,	
An ultraviolet flare at the centre of the elliptical galaxy NGC4552.
Nature, {\bf 378}, 39--41 (1995).

\noindent  \hangindent2em  \hangafter=1  5.\ Bade, N., Komossa, S., Dahlem, M.,	
Detection of an extremely soft X--ray outburst in the HII-like nucleus of NGC 5905.
Astronomy and Astrophysics, {\bf 309}, L35--L38 (1996).

\noindent \hangindent2em  \hangafter=1  6.\ Bloom, J. S., {\it et al.},
A possible relativistic jetted outburst from a massive black hole fed by a tidally disrupted star.
Science, {\bf 333}, 203--206 (2011).

\noindent  \hangindent2em  \hangafter=1  7.\ Sheeley Jr., N. R.,  Howard, R. A., Koomen, M. J., Michels, D. J., 
Coronagraphic observations of two new sungrazing comets.
Nature, {\bf 300}, 239--242 (1982).

\noindent  \hangindent2em  \hangafter=1  8.\ Harrington, J.,  LeBeau Jr., R. P.,  Backes, K. A., Dowling, T. E.,
Dynamic response of Jupiter's atmosphere to the impact of comet Shoemaker-Levy 9.
Nature, {\bf 368}, 525--527 (1994).

\noindent  \hangindent2em  \hangafter=1  9.\ Harwit, M., Salpeter, E. E.,
Radiation from comets near neutron stars.
Astrophysical Journal, {\bf 186}, L37--L39 (1973).

\noindent  \hangindent2em  \hangafter=1  10.\ Racusin, J. L., Cummings, J., Holland, S., Krimm, H., Oates, S. R., Page, K. L., Siegel, M.,	
{\it Swift} Observation of GRB 101225A.
GCN Report, 314 (2011).

\noindent    \hangindent2em  \hangafter=1  11.\ Th\"one, C. C., {\it et al.},
GRB 101225A shining the way to a new type of thermally dominated stellar explosion.
{\it Nature, this volume, arXiv:1105.3015 (2011).}

\noindent  \hangindent2em  \hangafter=1  12.\ Gehrels, N., Ramirez-Ruiz, E., Fox, D. B.,
Gamma-Ray Bursts in the Swift Era.
Annual Review Astronomy \& Astrophysics, {\bf 47}, 567--617 (2009).

\noindent  \hangindent2em  \hangafter=1  13.\ Campana, S., {\it et al.},
The association of GRB060218 with a supernova and the evolution of the shock wave.
Nature, {\bf 442}, 1008--1010 (2006).

\noindent  \hangindent2em  \hangafter=1  14.\ Chornock, R., Marion, G. H., Narayan, G., Berger, E., Soderberg, A. M., 
GRB 101225A: MMT spectroscopy.
GCN 11507 (2010).

\noindent  \hangindent2em  \hangafter=1  15.\ Tanvir, N. R., {\it et al.},
GRB 101225A: HST observations - no host detected.
GCN 11564 (2011).

\noindent \hangindent2em  \hangafter=1  16.\ G\"ansicke, B. T., {\it et al.},
A gaseous metal disk around a white dwarf.
Science, {\bf 314}, 1908--1910 (2006).

\noindent  \hangindent2em  \hangafter=1  17.\ Jura, M., 
A tidally disrupted asteroid around the white dwarf G29-38.
Astrophysical Journal, {\bf 584}, L91--L94 (2003).

\noindent  \hangindent2em  \hangafter=1  18.\ Ulmer, A.,
Flares from the Tidal Disruption of Stars by Massive Black Holes.
Astrophysical Journal, {\bf 514}, 180--187 (1999).

\noindent \hangindent2em  \hangafter=1  19.\ Cannizzo, J. K., Lee, H. M., Goodman, J.,
The disk accretion of a tidally disrupted star onto a massive black hole.
Astrophysical Journal, {\bf 351}, 38--46 (1990).

\noindent \hangindent2em  \hangafter=1  20.\ Burrows, D. N., {\it et al.},
Onset of a relativistic jet from the tidal disruption of a star by a massive black hole.
Nature, {\bf 476}, 421--424 (2011).

\noindent  \hangindent2em  \hangafter=1  21.\ Lodato, G., Rossi, E. M.,	
Multiband light curves of tidal disruption events.
Monthly Notices of the Royal Astronomical Society, {\bf 410}, 359--367 (2011).

\noindent \hangindent2em  \hangafter=1  22.\  Guillochon, J., Ramirez-Ruiz, E., Lin, D.,
Consequences of the ejection and disruption of giant planets.
Astrophysical Journal, {\bf 732}, 74 (2011).

\noindent \hangindent2em  \hangafter=1  23.\ Lasota, J.-P., 
The disc instability model of dwarf novae and low-mass X--ray binary transients.
New Astron. Rev., {\bf 45}, 449--508 (2001).

\noindent \hangindent2em  \hangafter=1  24.\ Lasota, J.-P., Dubus, G., Kruk, K., 
Stability of helium accretion discs in ultracompact binaries.
Astronomy and Astrophysics, {\bf 486}, 523--528 (2008).

\noindent  \hangindent2em  \hangafter=1  25.\ Shull, J. M., Stern, S. A.,	
Gamma-ray burst constraints on the galactic frequency of extrasolar Oort Clouds.
Astronomical Journal, {\bf 109}, 690--697  (1995).

\noindent \hangindent2em  \hangafter=1  26.\ Jura, M.,
An upper bound to the space density of interstellar comets.
Astronomical Journal, {\bf 141}, 155 (2011).

\noindent  \hangindent2em  \hangafter=1  27.\ Wolszczan, A., Frail, D. A.,
A planetary system around the millisecond pulsar PSR1257+12.
Nature, {\bf 355}, 145--147 (1992).

\noindent  \hangindent2em  \hangafter=1  28.\ Sigurdsson, S., Richer, H. B., Hansen, B. M., Stairs, I. H., Thorsett, S. E.,
A young white dwarf companion to pulsar B1620--26: evidence for early planet formation.
Science, {\bf 301}, 193--196 (2003).

\newpage

\noindent
{\bf Supplementary Information} is linked to the online version of the paper
at www.nature.com/nature.

\bigskip

\noindent
{\bf Acknowledgements} The authors acknowledge support from ASI and INAF. 
S. Campana wants to thank Andrea Possenti and Nial Tanvir for conversations, and
acknowledges Norbert Schartel for granting a DDT XMM-Newton observation.

\bigskip

\noindent
{\bf Author contributions} S. Campana led the research and wrote the majority of manuscript, dreamed up the tidal disruption model,
and did the X--ray data analysis. G.L. and E.M.R. worked on the tidal disruption 
model, fitted the light curves providing the minor body parameters, and estimated the rate of events. 
P.D'A. analysed optical data. N.P., M.D.V., G.T. and S. Covino contributed to the exclusion of Galactic models. G. Ghisellini, 
G. Ghirlanda, N.P., E.Pian  and M.D.V. contributed to the exclusion of extragalactic models. R.S. contributed to exclude the tidal disruption onto an 
intermediate mass black hole.
L.A.A., A.M., G.C., V.D'E., D.F., E. Palazzi, B.S. and S.D.V. contributed in obtaining optical data and provided an unbiased reading of the manuscript.

\bigskip

\noindent
{\bf Author Information} Reprints and permissions information is available at www.nature.com/reprints. 
The authors declare no competing financial interests. Readers are welcome to comment on the online 
version of this article at www.nature.com/nature. Correspondence and requests for materials should be 
addressed to S. Campana (sergio.campana@brera.inaf.it).

\newpage 

\begin{figure*}
        \parbox{3.5in}{
        \includegraphics[width=5in,angle=-90]{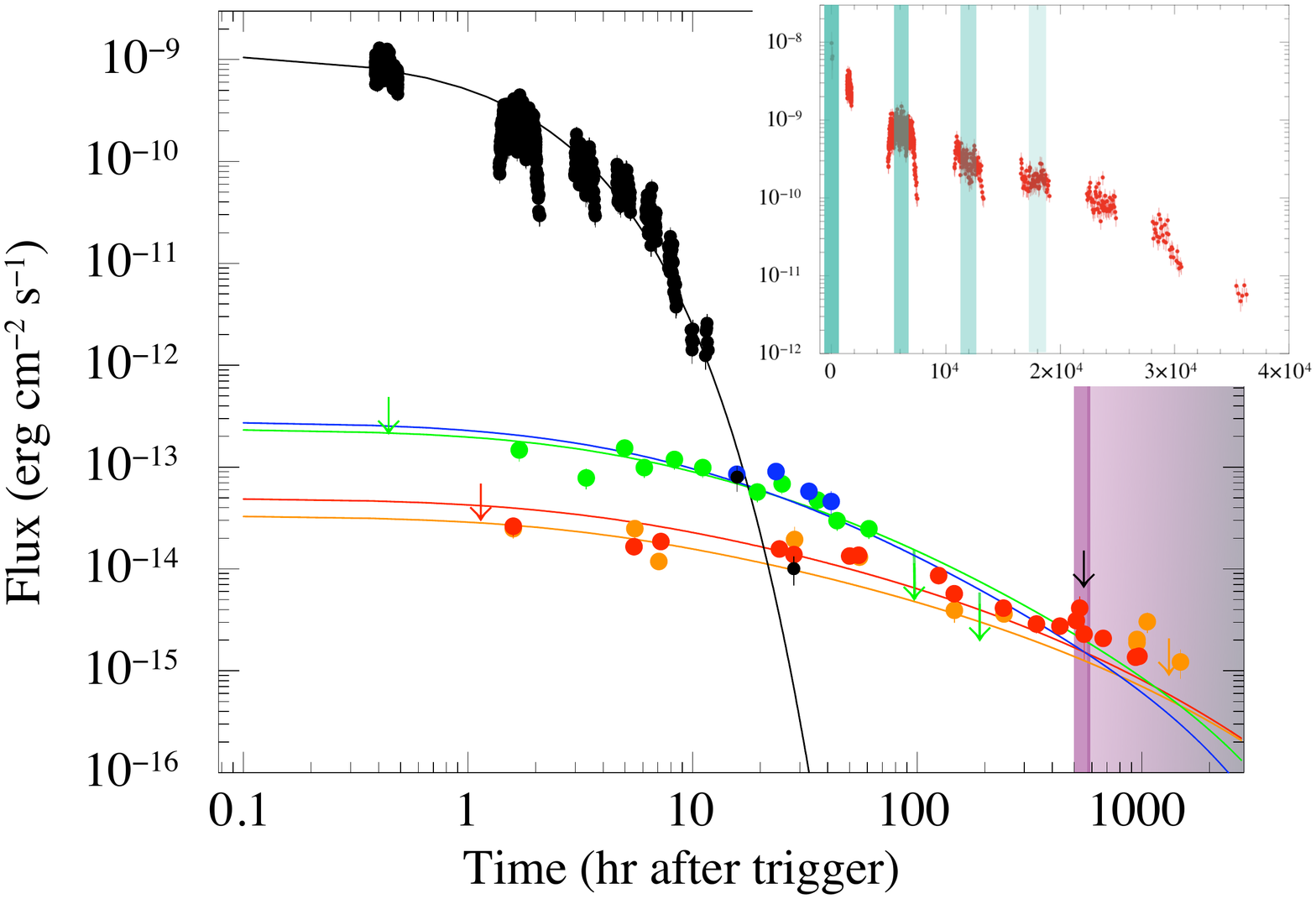}
\vskip -0.5truecm
        \centering{
\caption{\textbf{\bf Light curves of GRB 101225A.} 
GRB 101225A light curve in five energy bands: 
X--rays at 1 keV (black dots), UV at 2,030\,\AA\ (green) and 2,634\,\AA\ (blue), and optical at 6,400\,\AA\ 
($R$ band, red) and 7,700\,\AA\ ($I$ band, orange). Error bars, $1\,\sigma$.
The black arrow indicated the XMM-Newton upper limit, other arrows optical upper limits.
The X--ray light curve only refers to the disk contribution to the total flux ($\sim 0.3$ of the total, as derived from 
spectral modelling) and is corrected for the interstellar absorption
column density, of $N_H=(2.0\pm0.1)\times 10^{21}\cmdue$ (which is greater that the Galactic 
value,  $N_H^{\rm Gal}=7.9\times10^{20}$ cm$^{-2}$).
This enhanced column density may be due to the fraction of the minor body mass
that has been expelled from the system.
The continuous lines of different colors (the same as the data) represent the fit to the light curves using the 
phenomenological model of tidal disruption$^{21}$.
Because the model predicts a late transition of the accretion disk to a `cold' solution, the fit has been carried 
out up to $\sim 20$ d (the excluded region is indicated in pink-gray). A thick pink line indicates the time of the transition. 
Our model has four parameters: the mass of the minor body ($M_*$, 
we assume for simplicity a density of 1 g cm$^{-3}$, thereby fixing the radius, $R_*$), 
the periastron ($r_{\rm p}$), a geometrical factor ($D^2/\cos i$, where $D$ is the source distance and $i$ the source inclination) 
and the optical absorption ($A_V$). 
The best fit is obtained for: $M_*\sim5\times 10^{20}$ g,
$r_{\rm p}\sim9\times 10^3$ km, $D/\sqrt{\cos i}\sim3 \mbox{\ kpc}$ and 
$A_V\sim 0.8$ (in excess of the Galactic $A_V^{\rm Gal}=0.3$, consistent with the value determined by X--ray analysis).
The peak mass accretion rate with these parameters is 
$\dot{M}\approx 2\times 10^{16}$ g s$^{-1}$ 
and the peak luminosity $L\approx 3 \times 10^{36}$ erg s$^{-1}$, consistent with our 
hypothesis of a sub-Eddington accretion. 
In this regime, no emission lines are expected, at variance with super-Eddington events. 
{\it Inset:} X--ray light curve as observed by the Swift X--Ray Telescope.  
Shaded regions highlight the periastron passages calculated using our tidal disruption model, which have a timescale of 6,000--10,000 s. 
Outside each band, the X--ray emission markedly decreases, and this effect is strongest for the first passage. We verified that this 
effect is not instrumental in nature nor due to spectral changes. Inhomogeneities in the return debris are expected to give
rise to variability on the fall-back timescale.
}}}
\end{figure*}

\end{document}


\title{A Christmas minor body falling onto a neutron star: Supplementary Information}

\maketitle

\section*{Supplementary Methods}

\subsection*{X--ray data analysis}

\paragraph{{\it Swift} XRT \& UVOT data analysis}

The X-ray Telescope (XRT$^{28}$)
began observing GRB 101225A
on 2010 December 25 at 9:00:48.5 UT, 1383 seconds after the BAT trigger$^{29}$,
and ended on 2011 January 04 at 12:20:38 UT, collecting 3 ks in Windowed Timing (WT) mode, and
120 ks in Photon Counting (PC) mode, spread over 11 days. Supplementary Table \ref{xrtlog} reports the log of the XRT observations.

\begin{table}
\centering
\caption{{\it Swift} XRT observing log.}
\begin{tabular}{cccc}
\hline
Obs.ID.             &Start time                     & End time                            & XRT exp. time (s)\\
\hline
00441015000&2010-12-25 UT18:21:55&2010-12-26 UT03:01:15	&12527\\
00441015001&2010-12-26 UT02:57:31&2010-12-27 UT23:47:25	&33093\\
00441015003&2010-12-28 UT01:06:00&2010-12-28 UT16:37:09	&9694\\
00441015004&2010-12-29 UT01:08:01&2010-12-29 UT16:41:51	&10895\\
00441015005&2010-12-30 UT01:08:01&2010-12-30 UT16:40:26	&11033\\
00441015006&2010-12-31 UT02:49:01&2010-12-31 UT23:00:15	&9576\\
00441015007&2011-01-01 UT01:27:01&2011-01-02 UT00:41:47	&9949\\
00441015008&2011-01-02 UT01:28:01&2011-01-02 UT23:25:17	&11137\\
00441015009&2011-01-03 UT00:04:32&2011-01-03 UT23:30:15	&10853\\
00441015010&2011-01-04 UT06:39:01&2011-01-04 UT12:20:38	&4172\\
\hline
\end{tabular}
\label{xrtlog}
\end{table}

The GRB 101225A light curve at 1 keV shown in Fig. 1 of the main text has been obtained with the 
{\it Swift} burst-analyser provided by the UK {\it Swift} Science Data Centre at the University of Leicester$^{30}$. 
The original curve, already corrected for absorption, has been multiplied 
by a factor of $\sim 0.3$ to obtain the flux at 1 keV from the flux in the 0.3--10 keV energy band, obtained 
from spectral modeling (see below).

We extracted the XRT spectra of the first five orbits covered by {\it Swift} for spectral energy distribution (SED)
purposes. The XRT data were processed with standard procedures (XRTPIPELINE v0.12.6), filtering 
and screening criteria by using FTOOLS in the HEASOFT package (v.6.10). Both WT (first two orbits) and 
PC events were considered. The selection of event grades was 0-2 and 0-12 for WT and PC
data, respectively$^{28}$. 
We corrected for pile-up when required. The spectra were also corrected for point spread
function losses and vignetting generating the ancillary
response files with {\tt XRTMKARF}. We used the latest spectral redistribution matrices (v.012).
The XRT analysis was performed in the 0.3--10 keV energy band. 
Data from the first two orbits were taken in WT mode and spectra were rebinned
to have at least 50 counts per energy bin. Data from the last three orbits were 
obtained in PC mode and were rebinned in order to have 20 counts per energy bin.

The Ultraviolet/Optical Telescope (UVOT$^{31}$)
observed the target simultaneously with the XRT. A first account of these observations can be found in ref. 32.
We extracted two reference light curves in the $UVW1$ and $UVW2$ bands.
The data analysis was performed using the {\tt UVOTIMSUM} and {\tt UVOTSOURCE} tasks 
included in the HEASOFT. The latter task calculates the magnitude through aperture
photometry within a circular region and applies specific corrections
due to the detector characteristics. The reported magnitudes are in
the UVOT photometric system$^{33}$.

We also extracted five SEDs coincident with the first five orbits
of the {\it Swift} satellite. For the first two orbits data in all filters were available ($UVW2$, $UWM2$,
$UVM1$, $U$, $B$ and $V$), while for the remaining three only $V$, $B$ and $UVW2$ were present.
We summed different exposures in each orbit, and extracted a spectrum with the task {\tt UVOT2PHA}
for each filter and orbit. The extraction region was chosen to be 6 arcsec radius for UV filters and 3 arcsec
radius for optical filters due to a contaminating source. A separated closeby background region 
with 14 arcsec radius was adopted.

\paragraph{Early time spectral energy distribution}

We fit together the five data sets within the spectral energy package XSPEC.
The spectra cannot be fit with any single (absorbed) component model. We thus consider
a two-component model (see Supplementary Table \ref{xrtmodel}). For all the SEDs we keep tied the absorbing column density (modeled
with {\tt TBABS}$^{35}$)
to the same value. The Galactic column density is $N_H^{\rm Gal}=7.9\times 10^{20}\cmdue$ (ref. 36)
and the galactic extinction is  $A_V^{\rm Gal}\sim 0.3$ (ref. 37).
 A model made by a black body ({\tt BBODYRAD}) 
emission plus the emission from a multicolor accretion disk ({\tt DISKBB}) provides the best fit to the data. Alternatively, also 
the neutron star atmosphere ({\tt NSATMOS}$^{38}$)
model plus a disk black body emission provide a comparable good fit. 

\begin{table}
\begin{center}
\caption{{\it Swift} XRT and UVOT SED model fitting.}
\begin{tabular}{ccc}
\hline
Model$^*$   & Red. $\chi^2$ (dof) & N.h.p.$^+$\\
\hline
BB+DiskBB & 1.15  (848)      & $2\times 10^{-3}$ \\  
BB+PL         & 1.69    (848)    & $2\times 10^{-32}$\\   
BB+Brems. & 2.82 (848)        & $1\times 10^{-146}$\\ 
DiskBB+PL  & 1.36  (848)      & $2\times 10^{-11}$\\ 
NSA+DiskBB & 1.15  (852)    & $2\times 10^{-3}$\\  
NSA+PL         & 1.45  (852)     & $2\times 10^{16}$\\  
DiskBB+COMPTT$^{\S}$ & 1.04 (842) & $0.17$\\
\hline
\end{tabular}
\label{xrtmodel}
\end{center}

\medskip

\noindent $^*$ Models adopted in the fit of the five SEDs described in the text.
The model consists of two components (BB= black body; DiskBB= emission from an 
accretion disk; PL= Power Law; Brems.= bremsstrahlung emission; NSA= Neutron Star
Atmosphere with {\tt nsatmos} spectral model). All the double components model include absorption 
(using the model {\tt TBABS}). Spectral fits were carried out using the XSPEC package.

\noindent $^+$ Null hypothesis probability.

\noindent $^{\S}$ The normalisation $N$ of the disk black body component is relatively small.
The relation between the disk inner radius $r_{\rm in}$ and $N$ should be
$r_{\rm in}\times \sqrt{\cos {i}} = \sqrt{N} \times D_{10}$ km (where $i$ is the disk 
inclination and $D_{10}$ the source distance in 10 kpc units).
However,  the correct prescription to transform the normalization into
the inner disk radius depends on the effective temperature to color temperature ratio 
(which is basically unknown in our case$^{34}$).
In any case the relatively low value of the normalization calls for a relatively high disk inclination.

\end{table}

These two models can be easily interpreted in the present framework as emission from the accretion disk and 
from the neutron star surface  heated by the accreting matter. The black body model and the neutron star atmosphere 
models can be regarded as two extremes of the emission from the neutron star surface.
The best fit model is made by a disk component plus a black body component (or atmosphere component).
The presence of this latter component requires the presence of a hard surface. This clearly rules out black holes.

In the framework of accretion onto a neutron star surface, the black body or the neutron star atmosphere models 
represent a first attempt to model the physics of the process.
Matter accreting from the disk needs to reach the more slowly rotating neutron star.
This happens in the so-called boundary layer where a  large fraction of the accretion energy is released.
The boundary layer region is quite complex. The gas can reach very high temperatures, so Comptonization 
of soft  photons is (likely) the dominant energy loss mechanism$^{39,40}$.
 Boundary layer spectra are fit with Comptonization models,
with {\tt COMPTT} (within XSPEC) being the usual choice$^{41}$. 
We fit the same SEDs with a model made by a multicolor accretion disk ({\tt DISKBB}) plus a Comptonization 
model ({\tt COMPTT}), and an absorption component ({\tt TBABS}). The {\tt DISKBB} normalization is tied together 
in all the SEDs, depending on the disk inner radius. 
The fit is definitely better than the one with the black body plus accretion disk, with a reduced $\chi^2_{\rm red}=1.04$ 
(842 degrees of freedom, d.o.f.), although the disk parameters as well as the overall luminosities are similar.
Spectral parameters are reported in Supplementary Table \ref{comptt} and in Supplementary Figure 1.
The source unabsorbed luminosity in the 0.3--10 keV energy range is $1.7\times10^{36}$, $4.3\times10^{35}$, 
$1.9\times10^{35}$,  $1.1\times10^{34}$ and $5.9\times10^{34}$ erg s$^{-1}$, respectively. The bolometric 
corrections estimated over the 0.0001--1000 keV interval are 2.2, 1.8, 2.3, 2.3 and 1.6, respectively.

\begin{table}
\begin{center}
\caption{Spectral parameters of the Componization plus multicolor accretion disk models of the five SEDs.}
\begin{tabular}{ccccc}

\hline
Mid. time  & Disk temperature &  Seed Temperature & Plasma Temperature & Optical depth\\
(hr)            &  (keV)                     & (eV)                           & (keV)                             & \\
\hline
0.43 &      $1.81^{+0.16}_{-0.18}$ & $49^{+16}_{-23}$&$86^{+53}_{-21}$ & $0.38^{+0.21}_{-0.14}$\\
1.70 &      $1.00^{+0.10}_{-0.08}$ & $17^{+1}_{-3}$     &$56^{+1}_{-1}$      & $0.50^{+0.17}_{-0.13}$\\
3.34 &      $<0.59$                            & $14^{+1}_{-2}$     &$61^{+24}_{-26}$ & $0.51^{+0.20}_{-0.11}$\\
4.95 &      $<0.53$                            & $10^{+1}_{-1}$     &$53^{+11}_{-41}$ & $0.63^{+0.30}_{-0.11}$\\
6.55 &      $<0.48$                            & $11^{+1}_{-1}$     &$10^{+2}_{-3}$ & $<2.08$\\
\hline
\end{tabular}
\end{center}

\noindent The column density is $N_H=(2.0\pm0.1)\times10^{21}$ cm$^{-2}$.

\noindent
All the errors are $90\%$ confidence level for one parameters of interest, i.e. have been computed for
a $\Delta\chi^2=2.71$.

\label{comptt}
\end{table}

\begin{figure}
       \centering
        \parbox{6.5in}{
        \includegraphics[width=4in,angle=-90]{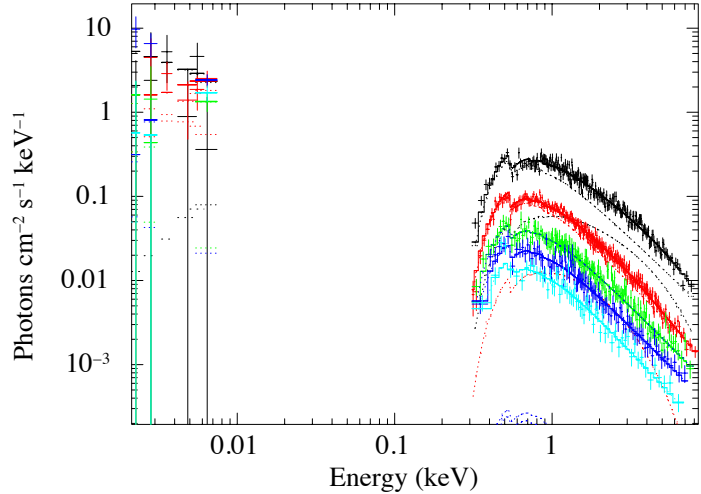}
        \centering\parbox{6in}{\vskip -1truecm
\caption{\label{fig:light2}{\bf Evolution of the spectral energy distribution with time.}
{\it Swift} XRT and UVOT unfolded spectra taken during the first five orbits were used to build  
spectral energy distributions (SED). The mid-times of the SEDs are 0.43 (black), 1.70 (red), 
3.34 (green), 4.95 (blue), 6.55 (magenta) hr, respectively. Here we show these five SEDs based on the 
best fit model made by a multicolor disk black body and a boundary layer emission model.
Their contribution is split and indicated with dotted lines.
This model provides a fair description of all the data with a reduced $\chi^2_{\rm red}=1.04$ 
(842 degrees of freedom, d.o.f.).
On the contrary, a black body plus a power law component model provides a reduced
$\chi^2_{\rm red}=1.69$ (848  d.o.f) and a disk black body plus a power law component
model  provides  a reduced $\chi^2_{\rm red}=1.36$ (848 d.o.f). These are firmly ruled out.
Given the upper limit derived for the disk temperature in the last three SEDs, the disk 
component shown here is just for the best fit temperature and does not reflect the maximum disk
component that it is possible to include in the data. In the first two spectra the multicolor disk 
black body component comprises $\sim 30\%$ and $\sim 10\%$ of the total flux.
The source unabsorbed luminosity in the 0.3--10 keV energy range is $1.7\times10^{36}$, 
$4.3\times10^{35}$, $1.9\times10^{35}$,  $1.1\times10^{34}$ and $5.9\times10^{34}$ erg s$^{-1}$, 
respectively. }}}
\end{figure}

\paragraph{Sharp drop of the X--ray flux} 

One of the key predictions of our model is that accretion of matter in the early stages 
of the minor body disruption is not steady and non-homogeneity in the retuning debris 
are expected during the first periastron passages. These give rise to variability on a fall-back timescale,
providing a quasi-periodicity (see e.g. Fig. 10 in ref. [22]).

Signs of this variability are easily visible in the X--ray light curve (see inset in Fig. 1).
We focus here on the data in the second orbit in WT mode (4940--7270 s). 
We positively check that the satellite attitude is stable and the source is far from bad columns
(whose presence is anyway accounted for and corrected).
We select the first 165 s (1272 counts) during the flux rise and the following 484 s (6286 counts).
We build the appropriate exposure maps and ancillary response files and fit the two spectra 
with the same absorbed power law with only the normalization free to vary. 
The fit provides an acceptable description of the data with a reduced $\chi^2_{\rm red}=1.00$ 
(with 274 d.o.f. and null hypothesis probability of $50\%$). 
This indicates that the variation is not induced by a spectral change, either an increase of the 
absorbing column density or a change in the power law photon index. This change can instead
be interpreted as variation in the mass accretion rate producing a flux variation of the same amount.

We repeat the same check on the late curve. We took 276 s (3325 counts) and 103 s (704 counts) 
before and after the flux decay. We repeat the same fit with an absorbed power law, finding a 
reduced $\chi^2_{\rm red}=1.06$ (with 159 d.o.f. and null hypothesis probability of $27\%$). 
Also in this case the variation is consistent with the hypothesis of not being induced by a spectral change.

\paragraph{{\it XMM-Newton} data analysis}

{\it XMM-Newton}$^{42}$
observed the field of GRB 101225A starting on  
Jan. 17, 2011 UT18:16:32  (23 d after the burst) for $\sim 32$ ks under a Directory discretionary 
Time program (PI Campana). 
All the three EPIC cameras were operated in Full Window mode with thin filters. 
We generated final product using {\tt SAS} v. 9.0.0, using the latest calibration files.
A high background interval affected the end of the observation resulting in 
23 ks of effective exposure for MOS1 and MOS2 cameras, and 21 ks for the pn camera.
No source is detected at the position of GRB 101225A in all images.
Assuming a worse-case power law spectrum with photon index $\Gamma=2$ and 
the Galactic column density, we derive a $3\,\sigma$ upper limit on the 0.5--10 keV
unabsorbed flux of $\sim 10^{-14}$ erg cm$^{-2}$ s$^{-1}$. A black body spectrum with 
temperature 0.3 keV, would result in a factor of $\sim 2$ tighter upper limit.

\subsection*{Optical/UV data analysis }

\paragraph{TNG data analysis}

We imaged the field of GRB 101225A with the Italian 3.6m Telescopio Nazionale Galileo (TNG), located in La Palma, 
Canary Islands. Optical $R$ and $I$-band observations were carried out with the Device Optimized for the LOw-Resolution 
(DOLoReS) camera on 2010 Dec. 31, 
2011 Jan. 16-17, Feb. 7 and Feb. 17-19. A $J$-band NIR observation was performed with the Near Infrared Camera Spectrometer 
(NICS)  on 2011 Feb. 8. All nights were clear, with seeing in the range $0.9''-1.2''$. The complete observing log 
is reported in Supplementary Table \ref{optlog}.
Image reduction was carried out following the standard procedures: subtraction of an averaged bias frame, division 
by a normalized flat frame. NIR frames were reduced using the jitter pipeline data reduction, part of the ECLIPSE 
package\footnote{http://www.eso.org/projects/aot/eclipse}. Astrometry of the optical and NIR images 
was performed using the 
USNOB1.0\footnote{http://www.nofs.navy.mil/data/fchpix} and 2MASS\footnote{http://www.ipach.caltech.edu/2mass} 
catalogues, respectively. Aperture photometry was made with the ESO-MIDAS\footnote{http://www.eso.org/projects/esomidas} 
{\tt DAOPHOT} task for all objects in the field. The photometric calibration was done against Landolt standard stars  
for optical images and against the 2MASS catalogue for NIR images. In order to minimize any systematic effect, we 
performed differential photometry with respect to a selection of local, isolated, and not-saturated reference stars visible in 
the field of view.

\begin{table}
\begin{center}
\caption{TNG Optical/NIR observing log.}
\begin{tabular}{ccccccc}
\hline
Time of obs.          &$\Delta_T$&Telescope           & Filter & Exposure        &Magnitude           &	    Flux\\
(UT)                        & (d)              &                              &           &          (s)            &    (Vega)              & ($\mu$Jy)\\
 \hline
20101231.86502 & 06.08880  &TNG (DOLoReS)&  $R$ &$3\times300$  &$23.75\pm0.13$&  $0.91\pm0.11$\\
20110116.84203 & 22.06581  &TNG (DOLoReS)&  $R$ &$7\times300$  &$24.10\pm0.29$&   $0.66\pm0.18$\\
20110117.85127 & 23.07505  &TNG (DOLoReS)&  $R$ &$12\times300$&$24.74\pm0.43$&   $0.37\pm0.15$\\
\hline
20101231.87893 & 06.10271  &TNG (DOLoReS)& $I$    &$3\times300$  &$23.36\pm0.26$&   $1.03\pm0.24$\\
20110207.87314 & 44.09692  &TNG (DOLoReS)&  $I$   &$20\times300$&$23.64\pm0.24$&   $0.80\pm 0.17$\\
20110217.86948 & 54.09326  &TNG (DOLoReS)& $I$    &$15\times120$&	 --                    &    -- \\
20110218.85349 & 55.07727  &TNG (DOLoReS)& $I$    &$20\times120$&	 --                    &   -- \\
20110219.84790 & 56.07168  &TNG (DOLoReS)& $I$    &$13\times180$&	 --                    &    -- \\ 
20110218.85696$^*$& 55.08074&TNG (DOLoReS)& $I$ &   6540             &$>24.5$ ($3\,\sigma$) & $<0.36$ ($3\,\sigma$)\\
\hline
20110208.86950 & 45.09328  &TNG (NICS)         &  $J$   &$56\times15$  &$>20.7$ ($3\,\sigma$)&   $<5.37$  ($3\,\sigma$)\\
\hline
\end{tabular}
\label{optlog}
\end{center}

\medskip
\noindent $^*$ Average of Feb. 17, 18 and 19 observation epochs.

\noindent Errors at the $1\,\sigma$ level. 

\end{table}

\paragraph{Optical light curve}

To build the light curve shown in the main text several observations were taken from 
Gamma-ray Burst Coordination Network (GCN) circulars.
In Supplementary Table \ref{gcn} we give a full account of them. Other observations were taken from Th\"one et al.$^{11}$,
reporting GCT observations. 

\begin{table}
\begin{center}
\caption{Optical observations from GCNs.}
\begin{tabular}{ccc|ccc}
\hline
Time of obs.       & Magnitude           & Refs. &Time of obs.       & Magnitude           & Refs.\\ 
(hr after $T_0$) &   (Vega)                & (hr after $T_0$)              & (Vega)                  & \\
 \hline
 $R$-band &  & & $I$-band &  & \\
1.14   &   $>21.5$         & 43 & & & \\
1.59   & $22.1\pm0.1$ &  44 &  1.59  & $21.5\pm0.2$ & 44\\
5.47   & $22.6\pm0.2$ &  45 & 5.47  & $21.6\pm0.2$ & 45\\
7.08   & $22.6\pm0.2$ &  46 & 7.07  & $22.3\pm0.2$ & 13 \\
7.22	  & $22.2\pm0.2$ & 13 & & & \\
24.2   & $22.2\pm0.2$  &  47& 24.2  & $21.6\pm0.2$ & 47\\
28.2   &  $22.8\pm0.2$ &  48 & 28.5  & $21.7\pm0.3$ & 13\\
50.0	  & $22.6\pm0.2$ & 13 & & & \\
55.2   &  $22.8\pm0.2$ &  49 & 55.2  & $22.3\pm0.2$ & 49\\
123    &  $23.3\pm0.2$ &  50 & & & \\
242    &  $24.1\pm0.1$ &  51 & 244   & $23.6\pm0.2$ & 13\\
338    &  $24.5\pm0.2$ &  52  & & & \\
432    &  $24.4\pm0.1$ & 14 & & & \\
512    & $24.2\pm0.1$ & 13& & & \\
674    & $24.6\pm0.1$ & 13& & & \\
939    & $25.1\pm0.1$ & 13& 952.3 & $24.3\pm0.2$ & 13\\
972    & $25.0\pm0.1$ & 13& 952.4 & $24.2\pm0.2$ & 13\\
1008  & $\gsim26.0$   &  53& 1486 & $24.8\pm0.3$ & 13 \\
\hline
 
\end{tabular}
\end{center}


\label{gcn}

\noindent Errors at the $1\,\sigma$ level. 

\end{table}

\subsection*{Modelling of the light curves: tidal disruption model}

We use the tidal disruption model of Lodato \& Rossi$^{21}$,
scaled down to a system comprised of a 
minor body object being disrupted by a neutron star. We also restrict the model to sub-Eddington luminosities, 
where the wind component -- considered in [21] 
-- is absent. 

We thus consider a minor body of mass $M_*$ and radius $R_*$, on a parabolic orbit around a 
neutron star of mass $M_{NS}=1.4\,M_{\odot}$ and radius $R_{NS}=12$ km, with periastron equal 
to $r_{\rm p}$. For simplicity, we assume that the minor body has a density $\rho= 1$ g cm$^{-3}$. 
If the periastron is smaller than the tidal radius $r_{\rm t}\sim 10^5-10^6$ km,
the minor body gets tidally disrupted. The presence of a magnetic field (if any) does not affect the 
infalling body until matter is fully ionized. Nearly half of the debris become unbound, while the other half are 
launched into highly eccentric orbits. We follow here the Rees'$^1$ 
suggestion, that assumes a flat 
distribution of mechanical energy in the debris. 
Taking into account the internal structure of the object may lead to some correction to this model at early times$^{54}$.
The most bound debris return to periastron after a time $t_{\rm min}$, given by:
\be
t_{\rm min}  = {{\pi}\over{2^{1/2}}} \left( {{r_{\rm p}\over{R_*}}} \right)^{3/2} \sqrt{{{r_{\rm p}^3}\over{G\,M_{NS}}}}.
\en
The other debris falls on a longer time depending on the specific energy $E$, with a fall-back time of 
\be
t_{\rm fb}={{2\,\pi\,G\,M_{NS}}\over({-2\,E})^{3/2}}
\en
Clearly the fall-back time has a strong dependence on the periastron radius (to the third power) and small changes in 
$r_{\rm p}$, determine large changes in $t_{\rm min}$ and $t_{\rm fb}$. Based on our model, we can estimate that $t_{\rm min}$
is in the 6,000-10,000 s range.

In the following, we identify $t_{\rm min}$ with the time of the BAT trigger. After $t_{\rm min}$, the rest of the debris 
return to periastron at a rate:
\be
\dot{M}(t) = {{1}\over{3}} {{M_*}\over{t_{\rm min}}} \left( {{t}\over {t_{\rm min}}} \right)^{-5/3} 
\en

The debris shock at periastron and rapidly circularize to form a thin accretion disk. The viscous timescale in the disk 
is initially shorter than the fall-back timescale and the disk can thus be approximated by a sequence of steady state models, 
extending from $R_{\rm in}\sim R_{NS}$ to the circularization radius $R_{\rm out}=2\,r_{\rm p}$. 
The effective temperature of the disk as a function of disk radius and time is given by 
\begin{equation}
\sigma T^4(R,t) = \frac{3\,G\,M_{NS}\dot{M}(t)}{8\pi R^3}\left(1-\sqrt{\frac{R_{\rm in}}{R}}\right).
\end{equation}
The emitted flux at frequency $\nu$ is computed assuming a multi-colour blackbody disk spectrum:
\begin{equation}
\nu F_{\nu}(t) = \frac{\cos i}{D^2}\int_{R_{\rm in}}^{R_{\rm out}}\frac{2\,h\,\nu^4}{c^2}\frac{2\pi R\mbox{d}R}{e^{h\,\nu/k\,T(R,t)}-1},
\end{equation}
where $D$ is the distance to the source and $i$ is the inclination of the disk with respect to the line of sight. 
We then redden the optical and UV fluxes, assuming the Cardelli$^{55}$ 
reddening law, with a visual extinction coefficient $A_V$, taken as a free parameter.

In total, our model has thus four free parameters: the mass of the minor body $M_*$, the periastron $r_{\rm p}$, 
the geometrical factor $D^2/\cos i$ and $A_V$. We find that the following set of 
parameters provides a good match to the X--ray, UV and optical light curves: $M_*=5\times 10^{20}$ g 
(which corresponds to a radius $R_*\approx 50$ km for the selected density), 
$r_{\rm p}=9\times 10^3$ km, $D/\sqrt{\cos i}=3 \mbox{ kpc}$
and $A_V=0.75$.
Note that the resulting periastron is well within the expected tidal radius for a minor body, 
and therefore consistent with the hypothesis that the object is tidally disrupted. The radius of the minor body 
is somewhat large but if the density is larger the radius will shrink. For a density of 10 g cm$^{-3}$ the radius 
will be $\approx 24$ km. Despite this the involved mass is not small in the framework of minor bodies.

The peak mass accretion rate with these parameters turns out to be 
$\dot{M}(t_{\rm min})\approx 2\times 10^{16}$ g s$^{-1}$ 
and the peak luminosity is thus $L(t_{\rm min})\approx 3 \times 10^{36}$ erg s$^{-1}$, consistent with our 
hypothesis of a sub-Eddington accretion. Note that this will imply that no emission lines are expected from super-Eddington 
wind as expected in tidal disruption models. 
Comparing this luminosity with the BAT peak flux (including a factor of 2.2 
for the bolometric correction based on spectral modeling) we obtain a distance of 
$\sim 3.5$ kpc.
We assumed a distance of 3 kpc, as motivated also by the matching of the 
observed fluence with the conversion of the minor body gravitational energy into radiation. The two estimates
match with very good approximation.
This distance places the neutron star in the Perseus arm. Its height above the Galactic plane is $\sim 1$ kpc.

\subsection*{Late time spectral energy distribution}

It is well known that accretion disks around compact objects are thermal-viscous unstable at a temperature 
corresponding to partial ionization of hydrogen. A model based on this instability captures the main properties 
of dwarf nova and transient low mass X--ray binary outbursts$^{22}$.
The same physics applies to disk made of helium or of heavier elements$^{23}$, 
leading to a critical mass inflow rate at a given radius,
where the instability sets in and then quickly propagates to the entire disk. 
In our case, the accretion rates decreases with time (Eq. 3), the disc cools down and eventually will hit the critical 
ionization temperature. The outer disc radius is the first radius at which this occurs. In our sub-Eddington regime, 
the outer disc is always gas pressure dominated, where the main opacity is Kramer (e.g. Fig 5.4 in ref. 56). 
Under these conditions, the temperature at outer disk $R_{\rm out}$ scales with the accretion rate as
\begin{equation}
T(R_{\rm out}) =  9.6\times 10^{3} (\dot{M}_{-13}(t))^{3/10} {\rm K},
\end{equation}
where $\mdot_{-13}$ is the mass accretion rate in units of $10^{-13}\msole$  yr$^{-1}$ and we assumed a 
standard Shakura \& Sunyaev disk and a viscosity parameter $\alpha=0.1$.
For low metallicity disc, the critical accretion rate is around $10^{-13}\msole$ yr$^{-1}$ (corresponding to transition 
temperature of $T \sim 10^4$ K, due partial hydrogen ionization$^{24}$).
 From Eqs. 3 and 6, we thus get that the the transition happens at $t \gsim 15$ d. 
It is well-known that the critical temperature decreases with metallicity$^{24}$.
Our disc will likely have solar metallicity or higher, so 15 d is just a lower limit.  If we take a slightly lower ionization 
temperature of $\sim 8,000$ K, we obtain a transition at $t\sim 20$ d.
We thus conservatively assume a transition time around 20 d. After the transition to the `cold' phase the disk emission would 
drop faster in the blue optical filters and slower in the red optical filters.

A second known effect occurs to the disk for decreasing mass inflow rates. For high mass inflow rate the mass inflow can reach 
the neutron star surface developing a boundary layer where matter is slowed down before accreting onto the neutron star surface.
If the neutron star possesses a significant magnetic field ($10^8-10^9$ G),  for a sufficiently low mass inflow rate this magnetic field will 
be able to disrupt the disk flow at a magnetospheric boundary. This happens when the magnetic pressure ($P_{\rm mag}(r) 
\propto B^4\,r^{-6}$, with $B$ the neutron star magnetic field) equates the disk ram pressure ($P_{\rm disk} (r) \propto \mdot\,r^{-5/2}$, 
where $\mdot$ is the mass accretion rate) at a radius larger than the neutron star radius$^{56}$.
When this occurs the disk inner edge ends at the magnetospheric boundary and for smaller radii the motion of the infalling matter
is controlled by the neutron star magnetic field. Since the magnetospheric radius  is proportional to $r_{\rm M}\propto B^{4/7}\,\mdot^{-2/7}$,
larger magnetospheric radii are expected for lower mass inflow rates. Since $\mdot$ decrease as $t^{-5/3}$, being a tidal disruption event, 
we can impose to our model that the inner radius of the disk $r_{\rm in}(t) \propto t^{-10/21}$. 
 
With these well-know ingredients we approached the late time SEDs. At very late times ($\gsim 20$ d), the mass accretion rate is very 
low and we expect a large magnetospheric radius (if the neutron star has a non-negligible magnetic field). Given that the outer disk is 
$\sim 2\,r_{\rm p}\sim 2\times10^9$ cm, we approximate the disk in this time interval as a small ring and treat its emission like a 
black body. We fit an optical-IR SED at 40 d taken from ref. 11. A black body fit (absorbed at $A_V\sim 0.75$ as 
derived from the light curve fitting) provides a good description of the data, with a reduced $\chi^2=1.7$ for 3 d.o.f. 
($16\%$ null hypothesis probability). The resulting temperature and radius are $T_{\rm BB}\sim 4500$ K and 
$R_{\rm BB}\sim 2.2\times 10^8$ cm (at 3 kpc, see Supplementary Figure 2). The derived temperature and emitting radius nicely fit with the
predictions of our late time disk model.

\begin{figure}[htb]
       \centering
        \parbox{6.5in}{
        \includegraphics[width=4in,angle=-90]{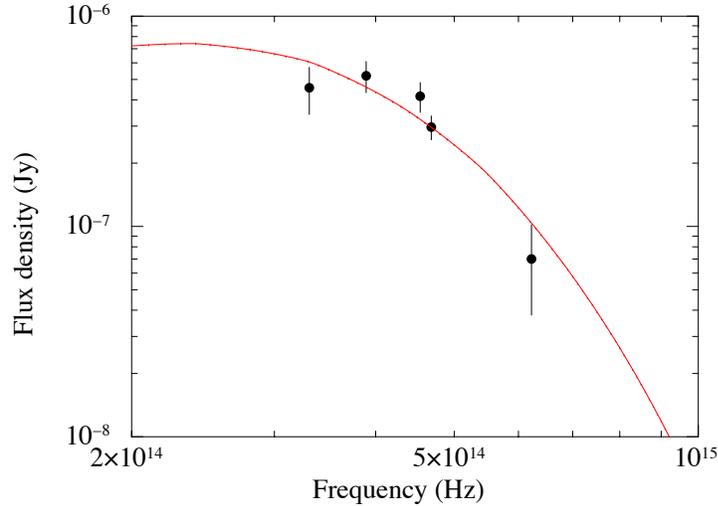}
        \centering\parbox{6in}{
          \caption{\label{fig:sed3}{\bf Late time ($\Delta t=40$ d) spectral energy distribution.} 
The late time spectral energy distribution taken from Th\"one et al.$^{11}$ 
has been fit with a absorbed ($A_V=0.75$ as derived from the light curve fitting) black body emission model.
The best fit model provides a temperature of 4500 K and a radius of $2.2\times 10^8$ cm (at 3 kpc).
The reduced $\chi^2_{\rm red}=1.7$ with 3 degrees of freedom, implies a null hypothesis probability of $16\%$.
}}}
\end{figure}  

Taking $R_{\rm BB}$ as the radius of the magnetosphere at 40 d, based on the model predictions about the evolution of the mass 
accretion rate we can estimate the magnetic field of the neutron star, which is $B\sim 10^9$ G. This is in line with the magnetic field 
of millisecond radio pulsars.

A very late time detection ($\Delta t=180$ d) has been reported$^{11}$. These data show a large decrease in the $r'$ band 
from data taken at $\Delta t=40$ d (factor of 4.6 in flux) but a mild decrease in the bluer $g'$ band (factor of $\sim 1.5$ in flux).
To model this spectrum we need a hot component that we identify with surface emission from the neutron star heated by the 
accretion episode and a cold component that may come from a dusty ring/disk, where the minor body formed.
Assuming a neutron star with a temperature of $\sim 10^6$ K we obtain a disk inner radius of $\sim 10^{13}$ cm and 
a temperature of $\sim 650$ K. Clearly with just two points this is just indicative\footnote{Assuming the same neutron star component to 
be present in the $\Delta t=40$ d data, the best fit parameters for the disk change only mildly to $R_{\rm BB}\sim 2.8\times 10^8$ 
cm and $T_{\rm BB}\sim 4000$ K}.

\subsection*{Why it cannot be another object?}

``When all possibilities have been eliminated, whatever remains, however improbable, must be the truth"
(Sir Arthur Conan Doyle).

%
%

\paragraph{Long-duration  Gamma--ray Burst}

GRB 101225A has been discovered as an image trigger. Inspection of the BAT light curve
indicates that there has not been a clear peak but a flat emission$^{10}$.
The event has also been detected by {\it MAXI}/GSC in the 2--10 keV energy band$^{57}$.
The burst is extremely long and the X--ray spectrum collected by the BAT is soft.
These peculiar characteristics resemble those of  GRB 060218$^{13}$
and GRB 100316D$^{58}$.
Also the UVOT flux is bright and similar to the shock break out emission seen in 
GRB 060218, but not in GRB 100316D. Moreover,
at variance with GRB 060218 and GRB 103016D, GRB 101225A does not show signs of a supernova (SN). 
This might not be an insurmountable problem since other two low energetic and close by ($z<0.13$) GRBs 
without SN have been observed in the past$^{59-61}$.
One of them, GRB 060614, has also a relative long duration ($T_{90}=102$ s, ref. 62),
whereas the other (GRB 060505) might be a short GRB. 
GRB 060614 had a quite spiky and hard X--ray emission (peak energy $E_{\rm p}>100$ keV) as detected by the BAT 
and it had a bright X--ray afterglow. In addition, GRB 060614 had a fast UV emission too but its emission declined 
achromatically$^{63}$.

All these bursts are very bright in XRT allowing us to disentangle the absorption pattern from our Galaxy (of a fixed amount) 
from that within the host galaxy, providing the redshift of the GRB$^{64}$
(see also S. Campana et al. 2012, in preparation).
The redshift of GRB 100316D has been predicted to be $0.014<z<0.28$ ($90\%$ confidence level$^{65}$),
in agreement with the observed value of $z=0.059$ (ref. 58). 
We applied the same technique to GRB 060614 providing $0.12<z<0.20$ (measured $z=0.125$) and to 
GRB 060218 providing $0.01<z<0.08$ (measured $z=0.033$).

Fitting the XRT X--ray spectrum of GRB 101225A with a Galactic plus intrinsic host galaxy absorption 
power law plus black body model led to the determination of the redshift  $z=0.07^{+0.13}_{-0.04}$ 
($90\%$ confidence level) with a reduced $\chi^2_{\rm}=1.07$ (505 d.o.f.$^{66}$).
Adding to these data also the optical/UV data described above and keeping fixed the 
Galactic column density to the measured value, the fit results to be highly unacceptable ($\chi^2_{\rm red}=2.01$, 847 d.o.f.).
A fit with a broken power law (instead of power law) with a difference in the power law indices of 0.5, as sometimes done 
when fitting such a large energy range based on GRB afterglow theory, results in an unacceptable fit too
($\chi^2_{\rm red}=1.45$, 842 d.o.f.). This clearly points against a GRB whose X--ray spectrum is always 
described a power-law.

Moreover, a low redshift solution has also to face the problem of a very faint optical galaxy.
A host galaxy candidate has been found in Gran Telescopio Canarias (GTC) data with $g'_{AB}\sim 27.2$ and 
$r'_{AB}\sim 26.9$ (ref. 11).
At a redshift of $z=0.07$ any host galaxy beneath 
the transient would have to be very faint ($M_g\sim -9$, i.e. fainter than 
a large globular cluster, e.g. $\omega$ Cen).
To make the host galaxy and the SN brighter one could think to move the object 
further away, but still within $z\lsim 1.1$ due to the detection in the $UVW2$ band. In this case 
however, the energetics of the GRB will become much larger and it will be a strong outlier of  the Amati relation$^{67}$.
To lie within $3\,\sigma$ from the Amati relation best fit, the maximum allowed redshift is $z\sim 0.3$. 
This would change the distance modulus by just $\sim 3.5$ mag, still leaving an extremely faint host galaxy
(see also below).

Given all these difficulties, we regard the GRB nature for GRB 101225A as highly unlikely.

\paragraph{Short-duration  Gamma--ray Burst}

A short duration GRB might mimic GRB 101225A if we are observing the so-called extended emission phase$^{62}$. 
The problem with this interpretation is that no hard X--ray spike has been observed 
associated with this event. Even if for some reason this spike has been overlooked, the extended 
emission observed in several short GRBs is too short compared (lasting only 100 to 1,000 s, ref. 68)
with what observed in GRB 101225A, and no bright UV long-duration emission 
has ever been detected in short GRBs. We therefore exclude this possibility.

\paragraph{Bright Supernova} 

Several cases of association between core-collapse SNe and GRBs have been discovered$^{69}$.
Given the large luminosity at maximum of these SNe, the association between GRB 101225A and a bright  SN is 
obviously ruled out by the lack of bumps in the optical light curve.
Assuming a SN template light curve and moving it at several redshifts (including
bolometric correction) we can exclude that a SN1998bw-like  and a 
SN2006aj-like are present in the data for $z\lsim 0.8$.

\paragraph{Faint Supernova} 

A different possibility is that extremely sub-luminous GRBs may be associated with ultra-faint 
Core-Collapse Supernovae$^{70,71}$.
Indeed, Th\"one et al.$^{11}$
suggested this possibility to explain the peculiarities of GRB 101225A.
They fit the late time SED with a SN1998bw template spectrum finding a good matching with a $\sim 10$ 
fainter supernova. This result has been obtained by varying contemporaneously the SN redshift (found to 
be $z=0.33^{+0.07}_{-0.04}$) and the stretching factor (which is determined for all the other SNe fixing the redshift to the 
known value).
We also note that all the other templates used (except two) provide statistically acceptable fits, spanning
the redshift range 0.2--0.5.

We fit the late light curve ($>100$ hr. i.e. $\sim 4$ d)  in the $R$ and $I$ bands (excluding the very late detection at $\sim 180$ d).
While in both light curves a small bump could be identified, a fit with a simple power law provides a statistically 
sound description of the data. In the $R$ band we obtain a reduced $\chi^2_{\rm red}=1.0$ (9 d.o.f., see Supplementary Figure 3) 
and in the $I$ band we obtain 
a reduced $\chi^2_{\rm red}=0.9$ (5 d.o.f.). Thus, the presence of a bump is not justified from a statistical point of view.
Note also that the bump, if present at all,  should peak at different times in the two bands: at $22\pm6$ d ($90\%$ confidence level) 
in the $R$ band and $\sim 48$ d in the $I$ band, respectively. This is due to the fact that in the $I$ band around $\sim 20$ d no
data are available, and in the $R$ band no data are available around $\sim 40$ d. This fact also testifies that the putative bump(s) 
are just statistical fluctuations.

A host detection has been claimed at $g'=27.2\pm0.3$ and $r'=26.9\pm0.14$ (ref. 11).
At a redshift of $z=0.33$ the absolute magnitude is $M_g=-13.7$, that is 2 mag fainter than the faintest host galaxy  
associated to a GRB-SN (GRB 060218). This host galaxy is much dimmer than the Magellanic Clouds and 
has the same absolute magnitude of the Fornax galaxy. It is peculiar that the dimmest supernova is also associated to 
the dimmest host galaxy.

\begin{figure}[ht]
       \centering
        \parbox{6.5in}{
        \includegraphics[width=4in,angle=-90]{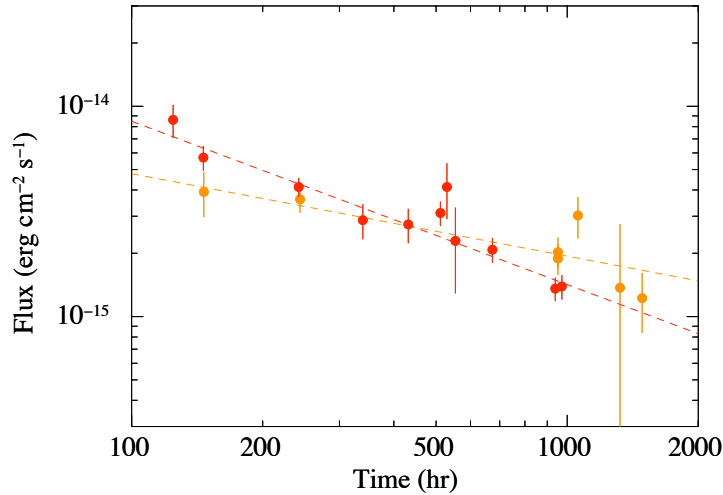}
        \centering\parbox{6in}{
          \caption{\label{fig:sed3}{\bf Late time $R$ and $I$ band light curves.} 
$R$ (red dots) and $I$ (orange dots) band curves after 100 hr from the trigger.
The dashed red line represents the best fit power law with index --0.8 to the $R$ band data; 
the dashed orange line the best fit (index --0.4) to the $I$ band data. The $R$ band slope is much steeper
than the $I$ band, indicating a chromatic behavior.
}}}

\end{figure}

\paragraph{Blazar/AGN}

Blazars show pronounced variability at all wavelengths. However, this variability is also accompanied 
by radio emission, that is lacking in this case. Radio emission has not been revealed with $3\,\sigma$ 
upper limit of  60 $\mu$Jy at 5 GHz, 4 d after the trigger$^{72}$,
and 42 $\mu$Jy and 30 $\mu$Jy at 4.5 and 7.9 GHz, 12 d after the trigger$^{73}$.
This excludes the blazar option.

\paragraph{Tidal disruption by Intermediate mass black hole}

The model used here$^{21}$ 
is self-similar and can provide the same results
scaling $M'_*=f^2\,M_*$, $M'_{NS}=f\,M_{NS}$, $D'=f\,D$, and $r'_{\rm p}=f^{1.8}\,r_{\rm p}$. 
This in principle can allow for an event occurring onto a supermassive black hole (requiring a 
star falling on it) or an intermediate mass black hole (IMBH, $\sim 1,000\msole$) in the local group
with an object like the Moon ($\sim 10^{25}$ g) falling on it.
A massive black hole is associated to a bright galaxy (based on the bulge luminosity - black hole mass 
correlation$^{74}$)
that is ruled out by the very faint quiescent counterpart. 
Moreover, all the observed events are characterized by much longer 
timescales (years$^{75,76}$), 
including the recent case of Swift J164449.3+373451$^{6,20}$.

The location of GRB 101225A lies somewhat close to the Andromeda galaxy (M31). If GRB 101225A belongs to this galaxy
it would be in its halo at a distance of at least $\sim 115$ kpc. In the distant local group it could also be associated to 
the dwarf spheroidal galaxy Andromeda XVIII ($\sim 1.4$ Mpc) at $\sim 12$ kpc from its center$^{53}$.
The IMBH possibility can be excluded based on the probability of having an IMBH well outside M31 
or in the outskirts of Andromeda XVIII, with a Moon falling on it. 

If the object is indeed extragalactic, the emitted energy is very large. Assuming a redshift $z=0.3$ the total emitted 
energy at high energies is $>10^{51}$ (ref. 11). This is very large and likely further excludes and extragalactic tidal event. 

\paragraph{X--ray binary transient}

Matter accreting onto a compact object in a binary system might be another option to explain GRB 101225A.
In this framework the luminosity is limited to be below the Eddington rate.
Assuming a black hole with mass $10\msole$ we derive a distance of $D\lsim75$ kpc.
This implies that the binary must be of Galactic origin. 
There is a large number of transient X--ray binaries. 
At 75 kpc, the late time detection from GTC would result in an 
absolute magnitude $M_R\sim 7.5$. This rules out all giants and supergiants companions 
(the absorption measured from the X--ray column density is low so it will not alter this conclusion).
The putative companion should be cooler than spectral type K2V.
This clearly points to a Low Mass X--ray Binary (LMXRB) system.
Outbursts of classical LMXRB transients have much longer timescales$^{77}$.
Shorter timescales 
are attained by relatively low luminosity transients (named faint transients$^{78}$)
linked to accreting 
millisecond X--ray pulsars. Binaries misidentified in the past as GRBs involve 
the so-called `burst-only' sources (sources discovered thanks to Type I X--ray burst activity
but with no or very weak persistent emission$^{79}$).
The X--ray spectrum during the outburst is usually characterized by a power law component 
that is not present in this case.
 
Lowering the burst peak to $10^{37}\ergs$ the maximum distance becomes $\sim 5$ kpc,
and the allowed companion should be cooler than spectral type M5V. This will also put strong 
limits on the orbital period of the system to ensure Roche lobe contact (the period should be 
less than a few hours). In such short orbital period binaries the companion star is always brighter
than the nominal main sequence star, due to irradiation from the primary.
Only ultracompact binaries do not show this effect ($P_{\rm orb}\lsim 1$ hr).
At 5 kpc the {\it XMM-Newton} upper limit implies a limiting luminosity of $\sim 3\times 10^{31}\ergs$.
All neutron star transients have a quiescent luminosity larger than this value$^{80}$
except one, the ultracompact binary candidate H1905+00 (ref. 81, 82)
out of $\sim 20$.
 
Black hole binaries have X--ray quiescent luminosities sometimes lower than the {\it XMM-Newton} limit,
but in this case the formation of a binary system with such a large mass ratio and small orbital period 
is not straightforward. Moreover, the X--ray spectral decomposition calls for a `surface' component besides
the accretion disk.

The ratio of the X--ray luminosity to the optical luminosity is $\sim 10^4$, which is about two orders of magnitude 
larger than observed in LMXRBs$^{83}$.
This is accounted for in our model by the small disk whereas in 
a LMXRB the disk is well developed from the truncation radius down to the compact object.

We also note that the {\it Swift} observations during the first two orbits showed a flux decrease at the end of each 
orbit, possibly suggesting a period of $\sim 2,500-3,000$ s. The presence of these dips is not confirmed in 
following observations. In addition, we carefully analyzed the X--ray spectra during these putative dips and
outside them during the first orbits. Absorption dips in LMXRBs are due to obscuration in
the thickened outer regions of an accretion disk$^{84}$.
A detailed spectral analysis of 
spectra (as well as a careful check of the satellite attitude) taken during the putative dips and outside 
did not reveal any change in the absorption column density, ruling out this possibility (see above).

On top of all these difficulties there is the position of GRB 101225A in the Galaxy. All X--ray binaries are concentrated
toward the Galactic center and bulge. The number of LMXRBs within $30\deg$ in latitude from the position of
GRB 101225A is 5 out of 185 binaries in the 4th Low-Mass X-Ray Binary Catalog$^{85}$.

Given all these constraints we regard the X--ray binary option too contrived.

\paragraph{Magnetar}

Magnetars are young ($\sim 10^6$ yr) isolated neutron stars with a high magnetic field (typically $B\gsim 10^{13}$ G) 
powered by magnetic energy rather than spin-down losses$^{86}$.
Magnetars emission is usually accompanied by repeated short ($<1$ s) high-energy flares that are not observed in 
GRB 101225A at any level. Pulsations are observed during outbursts with spin periods in the 2--12 s range. 
Bright quiescent X--ray counterparts are always observed  with $L_X\gsim 10^{35}\ergs$ due to the 
cooling of the hot neutron star and non-thermal emission associated with the huge magnetic field. 
None of these properties are shared with GRB 101225A.

\paragraph{Swift J195509.6+261406}

One possible related system is {\it Swift} J195509.6+261406, discovered as GRB 070610 by the {\it Swift} 
satellite. From this object X--ray and optical flares were observed, even if no pulsations were detected. 
Given the flaring variability and the peculiar afterglow with respect to GRBs it has been suggested that this 
source can be linked to magnetars$^{87-89}$.
A {\it Chandra} observation performed $\sim 2$ yr after the main event failed to detect the source in quiescence$^{80}$. 
The 0.3--10 keV unabsorbed flux limit is $\sim 10^{-15}\ergs\cmdue$. Tight upper limits 
were also derived in the optical and IR with $H>23$, $R>26.0$ and $i'>24.5$ (ref. 89,87).
Given this observational framework Rea et al.$^{80}$
questioned the magnetar interpretation, based on the
too low quiescent X--ray flux. They concluded that the source must lie in the Galactic halo to be minimally 
consistent with the magnetar scenario. Clearly this option is rather unlikely given the paucity of magnetars in our Galaxy.
Rea et al.$^{80}$ 
suggested instead an X--ray binary origin for {\it Swift} J195509.6+261406, even if the very unusual 
flaring behavior remains unexplained.

Here we note that more energetic and shorter events than GRB 101225A can occur in the tidal capture of a minor body
by a neutron star if the tensile strength of the minor body is strong. In this case one can observe a single, bright and short 
event if the object is accreted by the neutron star directly (as was suggested in the past to explain short GRBs$^{9,90,91}$.
Alternatively if the minor body is 
disrupted into several pieces still part of an accretion disk, one can observe a behavior similar to GRB 101225A with 
repeated flares on top. One possibility is therefore that {\it Swift} J195509.6+261406 is not a magnetar nor an X--ray binary 
but a tidal disruption event similar to GRB 101225A. 

\paragraph{Nova}

Recently the `Nova-like' system V407 Cyg has been detected by {\it Fermi}$^{92}$,
so in principle 
one can expect to detect an X--ray flare, associated with Nova-like systems.  However, this possibility is easily ruled out 
by the fact that Novae at maximum light are extremely luminous at optical wavelengths. For example, fast evolving Novae, 
on time scale of 1--2 d, have an absolute magnitude at maximum of $\sim -9$ (ref. 93),
that corresponds to $V\sim 11-15$ for novae in the local group of galaxies. These bright counterparts  have not been 
observed  in temporal and spatial coincidence with GRB 101225A.  Even if one assumes to have missed the maximum, after 
15 d a nova in the local group of galaxies should still have an apparent magnitude brighter than  $V \sim 18$ (ref. 94).
Also this case is excluded by optical observations. 
On average, recurrent Novae are fainter at maximum by 2--3 mag than classical Novae$^{95}$.
Current optical observations  rule out also this possibility.

\paragraph{Flare star}

Active stars can give rise to impulsive fast-rising (seconds-minutes)  X--ray flares that can last from a few minutes up to hours or days. 
These flares can be orders of magnitude more powerful than those observed in the Sun. The brightest flares have been observed in young 
stars like AB Dor$^{96}$,
RS CVn type of stars$^{97,98}$ 
and Algol type of stars$^{99}$.
Usually these flares are quite soft when compared to a GRB prompt emission, however in a few cases the emission 
has been detected up to 50 keV and beyond (e.g. for Algol$^{100}$,
UX Ari$^{97}$, 
and II Peg$^{98}$).
The energy budget observed for the most powerful of these events are in the range $10^{36}-10^{37}$ erg with a peak luminosity of a few 
$10^{32}$ erg s$^{-1}$. Less powerful (but still more powerful than in the Sun) flares are commonly observed in the UV Ceti-type flares stars, 
with an energy budget of $10^{30}-10^{34}$ erg and a peak luminosity of $10^{28}-10^{30}$ erg s$^{-1}$ (refs. 101, 102).
Still they have an X--ray to bolometric luminosity ratio as high as the above type of flare stars. To check if one of these 
type of flare stars could be responsible for the event observed by {\it Swift}, we divided the peak luminosity observed for the most powerful 
flares of these stars by the peak flux observed by the {\it Swift} XRT and derived the maximum distance to which these stars could have been 
placed (even more stringent values would be derived if we use the {\it Swift} BAT prompt peak flux). We derived a maximum distance of $\lsim 
50$ pc for the RS CVn type of stars and $\lsim 10$ pc of the UV Ceti type of stars. At these distances these stars should have an apparent 
$R$ magnitude of 7--9 for the RS CVn and of 10--18 (two magnitudes fainter in quiescence) for the UV Ceti. The faintness of the optical 
counterpart detected during the flare and in quiescence by the {\it Swift} UVOT and by optical telescopes (including HST) strongly argues 
against an association between a flare star and GRB 101225A.


\clearpage

{\bf \Large References}

\bigskip


\noindent 28.\ Burrows, D. N., {\it et al.}, 
The {\it Swift} X--Ray Telescope. 
Sp. Sci. Rev., {\bf 120}, 165--195 (2005).

\noindent 29.\ Racusin, J. L.,  {\it et al.}, 
Trigger 441015: {\it Swift} detection of a possible burst or transient.
GCN Circ. 11493 (2010).

\noindent   \hangindent2em  \hangafter=1
30.\  Evans, P. A., {\it et al.}, 
The {\it Swift} Burst Analyser. I. BAT and XRT spectral and flux evolution of gamma ray bursts.
Astronomy and Astrophysics, {\bf 519}, A102 (2010).

\noindent 31.\  Roming, P. W. A., {\it et al.}, 
The {\it Swift} Ultra-Violet/Optical Telescope. 
Sp. Sci. Rev. {\bf 120}, 95--142 (2005).

\noindent 32.\  Siegel, M. H., Racusin, J. L., GRB 101225A: {\it Swift}/UVOT Detection.
 GCN Circ. 11499 (2010).

\noindent  \hangindent2em  \hangafter=1 
33.\  Poole, T. S., {\it et al.}, Photometric calibration of the {\it Swift} ultraviolet/optical telescope.
Monthly Notices of the Royal Astronomical Society, {\bf 383}, 627--645 (2008).

\noindent 	 \hangindent2em  \hangafter=1
34.\  Ebisawa, K., \.Zycki, P., Kubota, A., Mizuno, T., Watarai, K.,	
Accretion disk spectra of ultraluminous X--ray sources in nearby spiral galaxies and galactic superluminal jet sources.
Astrophysical Journal, {\bf 597}, 780-797 (2003).

\noindent  \hangindent2em  \hangafter=1
35.\  Wilms, J., Allen, A., McCray, R.,	
On the absorption of X--rays in the interstellar medium.
Astrophysical Journal, {\bf 542}, 914--924 (2000).

\noindent  \hangindent2em  \hangafter=1
36.\  Kalberla, P. W. A., Burton, W. B., Hartmann, D., Arnal, E. M., Bajaja, E., Morras, R., P\"oppel, W. G. L., 
The Leiden/Argentine/Bonn (LAB) Survey of Galactic HI. Final data release of the combined LDS and IAR surveys with improved 
stray-radiation corrections. 
Astronomy and Astrophysics, {\bf 440}, 775--782 (2005).

\noindent  \hangindent2em  \hangafter=1 
37.\  Schlegel, D. J., Finkbeiner, D, P., Davis, M.,
Maps of dust infrared emission for use in estimation of reddening and cosmic microwave background radiation foregrounds.
Astrophysical Journal, {\bf 500}, 525--553 (1998).

\noindent  \hangindent2em  \hangafter=1
38.\  Heinke, C. O., Rybicki, G. B.,  Narayan, R., Grindlay, J. E., 
A hydrogen atmosphere spectral model applied to the neutron star X7 in the globular cluster 47 Tucanae.
Astrophysical Journal, {\bf 644}, 1090--1103 (2006).

\noindent  \hangindent2em  \hangafter=1 
39.\  Popham, R., Sunyaev, R.,
Accretion disk boundary layers around neutron stars: X--ray production in Low-Mass X--Ray Binaries.
Astrophysical Journal, {\bf 547}, 355--383 (2001).

\noindent  \hangindent2em  \hangafter=1 
40.\  Grebenev, S. A., Sunyaev, R. A.,
The formation of X--ray radiation in a boundary layer during disk accretion onto a neutron star.
Astronomy Letters, {\bf 28}, 150--162 (2002).

\noindent  \hangindent2em  \hangafter=1
41.\  Titarchuk, L.,
Generalized Comptonization models and application to the recent high-energy observations.
Astrophysical Journal, {\bf 434}, 570--586 (1994).

\noindent  \hangindent2em  \hangafter=1 
42.\  Jansen, F., {\it et al.},
XMM-Newton observatory. I. The spacecraft and operations.
Astronomy and Astrophysics, {\bf 365}, L1--L6 (2001).

\noindent  \hangindent2em  \hangafter=1
43.\  Andreev, M., Sergeev, A., Pozanenko, A., 
GRB 101225A (Swift trigger 441015): optical upper limit.
GCN 11494 (2010).

\noindent  \hangindent2em  \hangafter=1 
44.\  Xu, D., Ilyin, I.,  Fynbo, J. P. U.,
Trigger 441015 / GRB 101225A: NOT optical afterglow candidate.
GCN 11495 (2010a).

\noindent  \hangindent2em  \hangafter=1 
45.\  Xu, D., Ilyin, I.,  Fynbo, J. P. U.,
Trigger 441015 / GRB 101225A: Further NOT optical afterglow candidate.
GCN 11496 (2010b).

\noindent  \hangindent2em  \hangafter=1
46.\  Park, W.-K., Im, M., Choi, C., Jeong, H., Lim, J., Pak, S., 
GRB 101225A: CQUEAN optical observation.
GCN 11594 (2011).

\noindent 47.\  Th\"one, C. C., {\it et al.},
GRB 101225: Optical follow-up of the Christmas burst.
GCN 11503 (2010).

\noindent 48.\  Wiersema, K., Tanvir, N.,  Levan, A., 
GRB 101225A: WHT observations.
GCN 11502 (2010).

\noindent 49.\  Cenko S. B., 
GRB 101225A: P60 Observations.
GCN 11506 (2010).

\noindent 50.\  Xu, D., Hakala, P.,  Fynbo, J. P. U.,
GRB 101225A: optical break from NOT observation
GCN 11508 (2010c).

\noindent  \hangindent2em  \hangafter=1 
51.\  Xu, D., Malesani, D., Fynbo, J. P. U.,  Hjorth, J.,
Jakobsson, P.,  Augusteijn, T.,
GRB 101225A: Host galaxy.
GCN 11519 (2011).

\noindent  \hangindent2em  \hangafter=1
52.\  Fynbo, J. P. U., Xu, D., 
GRB 101225A: Redshift retraction and results of additional photometric follow-up at the NOT.
GCN 11563 (2011).

\noindent  53.\ Levan, A. J., Tanvir, N. R., 
GRB 101225A is likely at $z\sim 0$.
GCN 11642 (2011).

\noindent   \hangindent2em  \hangafter=1
54.\ Lodato, G., King, A. R., Pringle, J. E.,	
Stellar disruption by a supermassive black hole: is the light curve really proportional to $t^{-5/3}$?
Monthly Notices of the Royal Astronomical Society, {\bf 392}, 332--340 (2009).

\noindent  \hangindent2em  \hangafter=1
55.\  Cardelli, J. A., Clayton, G. C., Mathis, J. S.,	
The relationship between infrared, optical, and ultraviolet extinction.
Astrophysical Journal, {\bf 345}, 245--256 (1989).

\noindent 56.\ 	Frank, J., King, A., Raine, D. J.,
Accretion Power in Astrophysics (3rd Ed.).
Cambridge University Press (2002).

%
%

\noindent 57.\  Serino, M., {\it et al.},
GRB 101225A: MAXI/GSC observations.
GCN 11505 (2011).

\noindent  \hangindent2em  \hangafter=1
58.\  Starling, R. L. C., {\it et al.},
discovery of the nearby long, soft GRB 100316D with an associated supernova.
Monthly Notices Royal Astronomical Society, {\bf 411}, 2792--2803 (2011).

\noindent  \hangindent2em  \hangafter=1
59.\  Della Valle, M., {\it et al.}, 
An enigmatic long-lasting $\gamma-$ray burst not accompanied by a bright supernova.
Nature, {\bf 444}, 1050--1052 (2006).

\noindent  \hangindent2em  \hangafter=1
60.\  Fynbo, J. P. U., {\it et al.}, 	
No supernovae associated with two long-duration $\gamma-$ray bursts.
Nature, {\bf 444}, 1047--1049 (2006).

\noindent  \hangindent2em  \hangafter=1
61.\  Gal-Yam, A., {\it et al.}, 
A novel explosive process is required for the $\gamma-$ray burst GRB 060614.
Nature, {\bf 444}, 1053--1055 (2006).


\noindent 62.\  Gehrels, N., {\it et al.}
A new $\gamma-$ray burst classification scheme from GRB 060614.
Nature, {\bf 444}, 1044--1047 (2006).

\noindent  \hangindent2em  \hangafter=1 
63.\  Mangano, V., {\it et al.},
Swift observations of GRB 060614: an anomalous burst with a well behaved afterglow.
Astronomy and Astrophysics, {\bf 470}, 105--118 (2007).

\noindent  \hangindent2em  \hangafter=1
64.\  Campana, S., Th\"one, C. C., de Ugarte Postigo, A., Tagliaferri, G., Moretti, A., Covino, S.,
The X-ray absorbing column densities of Swift gamma-ray bursts.
Monthly Notices Royal Astronomical Society, {\bf 402}, 2429--2435 (2010).

\noindent 65.\  Campana, S.,
Tentative redshift of GRB100316D from X-ray data.
GCN 10571 (2010).

\noindent  \hangindent2em  \hangafter=1
66.\  Campana, S., Covino, S., Racusin, J. L., Page, K. L., 
Tentative redshift of GRB101225A (Christmas's burst) from Swift-XRT data
GCN 11501 (2010).

\noindent   \hangindent2em  \hangafter=1
67.\  Amati, L., Frontera, F., Guidorzi, C.,	
Extremely energetic Fermi gamma-ray bursts obey spectral energy correlations.
Astronomy and Astrophysics, {\bf 508}, 173--180 (2009).

\noindent  \hangindent2em  \hangafter=1
68.\  Sakamoto, T., Gehrels, N.,
Indication of two classes in the Swift short gamma-ray bursts from the XRT X-ray afterglow light curves.
in ``Sixth Huntsville Symposium''. AIP Conference Proc., {\bf 1133}, 112--114 (2009). 

\noindent  \hangindent2em  \hangafter=1
69.\  Woosley, S. E., Bloom, J. S.,
The Supernova Gamma---ray burst connection.
Annual Review of Astronomy \& Astrophysics, {\bf 44}, 507--556 (2006).

\noindent 70.\  Pastorello, A., {\it et al.},
A very faint core-collapse supernova in M85.
Nature, {\bf 449},  1--2 (2007).

\noindent  \hangindent2em  \hangafter=1 
71.\  Valenti, S., {\it et al.},  
A low-energy core-collapse supernova without a hydrogen envelope.
Nature, {\bf 459}, 674--677 (2009).

\noindent 72.\  Zauderer, A., Berger, E., Fong, W.,
GRB 101225A: early EVLA observation of Christmas Burst.
GCN 11770

\noindent 73.\  Frail, D. A., 
GRB 101225A.
GCN 11550

\noindent 	 \hangindent2em  \hangafter=1
74.\  Magorrian, J., {\it et al.},
The demography of massive dark objects in galaxy centers.
Astronomical Journal, {\bf 115}, 2285--2305 (1998).

\noindent  \hangindent2em  \hangafter=1 
75.\  Komossa, S.,
X--ray evidence for supermassive black holes at the centers of nearby, non-active galaxies.
Reviews in Modern Astronomy, {\bf 15}, 27--56 (2002).
	
\noindent  \hangindent2em  \hangafter=1 
76.\  Gezari, S.,  {\it et al.}, 
Ultraviolet detection of the tidal disruption of a star by a supermassive black hole.
Astrophysical Journal, {\bf 653}, L25--L28 (2006).

\noindent  \hangindent2em  \hangafter=1 
77.\ Campana, S., Colpi, M., Mereghetti, S., Stella, L., Tavani, M.,
The neutron stars of soft X--ray transients.
Astronomy and Astrophysics Review, {\bf 8}, 279--316 (1998).

\noindent  \hangindent2em  \hangafter=1 
78.\  Wijnands, R.,
Accretion-driven millisecond X--ray pulsars.
proc. of `Trends in Pulsar Research', J. A. Lowry ed. Nova Science Publishers, New York, 
53--78(2006).

\noindent  \hangindent2em  \hangafter=1
79.\  Campana, S., 
Linking burst-only X--ray binary sources to faint X--ray transients
Astrophysical Journal, {\bf 699}, 1144--1152 (2009).

\noindent  \hangindent2em  \hangafter=1 
80.\  Rea, N., Jonker, P. G., Nelemans, G., Pons, J. A., Kasliwal, M. M., Kulkarni, S. R., Wijnands, R.,
The X--ray quiescence of {\it Swift} J195509.6+261406 (GRB 070610): an optical bursting X--ray binary?
Astrophysical Journal, {\bf 729}, L21--L25 (2011).

\noindent  \hangindent2em  \hangafter=1
81.\  Jonker, P. G., Bassa, C. G., Nelemans, G., Juett, A. M., Brown, E. F., Chakrabarty, D.,	
The neutron star soft X--ray transient 1H 1905+000 in quiescence.
Monthly Notices of the Royal Astronomical Society, {\bf 368}, 1803--1810 (2006).

\noindent  \hangindent2em  \hangafter=1
82.\  Jonker, P. G., Steeghs, D., Chakrabarty, D., Juett, A. M.,	
The cold neutron star in the soft X--ray transient 1H 1905+000.
Astrophysical Journal, {\bf 665}, L147--L150 (2007).

\noindent  \hangindent2em  \hangafter=1 
83.\  Della Valle, M., Jarvis, B. J., West, R. M., 
Evidence for a black hole in the X--ray nova MUSCAE 1991.
Nature, {\bf 353}, 50--52 (1991).

\noindent  \hangindent2em  \hangafter=1
84.\  White, N. E., Swank, J. H.,	
The discovery of 50 minute periodic absorption events from 4U 1915-05.
Astrophysical Journal, {\bf 253}, L61--L66 (1982).

\noindent  \hangindent2em  \hangafter=1
85.\  Liu, Q. Z., van  Paradijs, J., van  den  Heuvel, E. P. J.,
A catalogue of low-mass X--ray binaries in the Galaxy, LMC, and SMC (Fourth edition).
Astronomy and Astrophysics, {\bf 469}, 807--810 (2007).

\noindent  \hangindent2em  \hangafter=1
86.\  Mereghetti, S., 
The strongest cosmic magnets: soft gamma-ray repeaters and anomalous X--ray pulsars.
Astronomy and Astrophysics Review, {\bf 15}, 225--287 (2008).

\noindent  \hangindent2em  \hangafter=1
87.\  Castro-Tirado, A. J., {\it et al.},
Flares from a candidate Galactic magnetar suggest a missing link to dim isolated neutron stars.
Nature, {\bf 455}, 506--509 (2008).

\noindent  \hangindent2em  \hangafter=1
88.\  Stefanescu, A., Kanbach, G., S\-lowikowska, A., Greiner, J., McBreen, S., Sala, G.,	
Very fast optical flaring from a possible new Galactic magnetar.
Nature, {\bf 455}, 503--505 (2008). 

\noindent  \hangindent2em  \hangafter=1
89.\  Kasliwal, M. M., {\it et al.}, 
GRB 070610: A Curious Galactic Transient.
Astrophysical Journal, {\bf 248}, 1127--1135 (2008).

\noindent 	 \hangindent2em  \hangafter=1
90.\  Newman, M. J., Cox, A. N.	,
Collisions of asteroids on neutron stars as a cause of cosmic gamma-ray bursts.
Astrophysical Journal, {\bf 242}, 319--325 (1980).


\noindent   \hangindent2em  \hangafter=1
91.\  Colgate, S. A., Petschek, A. G.,	
Gamma ray bursts and neutron star accretion of a solid body.
Astrophysical Journal, {\bf 248}, 771--782 (1981).

\noindent  \hangindent2em  \hangafter=1
92.\  Abdo, A. A., {\it et al.},
Gamma-ray emission concurrent with the nova in the symbiotic binary V407 Cyg.
Science {\bf 329}, 817--820 (2010).

\noindent  \hangindent2em  \hangafter=1
93.\  Della Valle,  M., Livio, M.,
The calibration of Novae as distance indicators.
Astrophysical Journal, {\bf 452}, 704-709 (1995).

\noindent  \hangindent2em  \hangafter=1
94.\  Capaccioli, M. , Della Valle, M., D'Onofrio, M., Rosino, L.,
Properties of the nova population in M31.
Astronomical Journal, {\bf 97}, 1622--1633 (1989).

\noindent  \hangindent2em  \hangafter=1
95.\  Rosino, L., Capaccioli, M., D'Onofrio, M., della Valle, M.,	
Fifty-two novae in M31 discovered and observed at Asiago from 1971 to 1986.
Astronomical Journal, {\bf 97}, 83--96 (1989).

\noindent  \hangindent2em  \hangafter=1
96.\  Maggio, A., Pallavicini, R., Reale, F., Tagliaferri, G.,
Twin X-ray flares and the active corona of AB Dor observed with BeppoSAX.
Astronomy and Astrophysics, {\bf 356}, 627--642 (2000).

\noindent  \hangindent2em  \hangafter=1
97.\  Franciosini, E., Pallavicini, R., Tagliaferri, G.,
BeppoSAX observation of a large long-duration X-ray flare from UX Arietis.
Astronomy and Astrophysics, {\bf 375}, 196--204 (2001).

\noindent  \hangindent2em  \hangafter=1
98.\  Osten, R. A., Drake, S., Tueller, J., Cummings, J., Perri, M., Moretti, A., Covino, S.,
Nonthermal hard X--ray emission and iron K$\alpha$ emission from a superflare on II Pegasi.
Astrophysical Journal, {\bf 654}, 1052--1067 (2007).

\noindent 	 \hangindent2em  \hangafter=1
99.\  Schmitt, J. H. M. M., Favata, F.,
Continuous heating of a giant X--ray flare on Algol.
Nature, {\bf 401}, 44--46 (1999).

\noindent  \hangindent2em  \hangafter=1
100.\  Favata, F., Schmitt, J. H. M. M.,
Spectroscopic analysis of a super-hot giant flare observed on Algol by BeppoSAX on 30 August 1997.
Astronomy and Astrophysics, {\bf 350}, 900--916 (1999).

\noindent  \hangindent2em  \hangafter=1
101.\  Pallavicini, R., Tagliaferri, G., Stella, L.,
X--ray emission from solar neighbourhood flare stars - A comprehensive survey of EXOSAT results.
Astronomy and Astrophysics, {\bf 228}, 403--425 (1990).

\noindent  \hangindent2em  \hangafter=1
102.\  Gershberg, R. E., Shakhovkay, N. I., 
Characteristics of activity energetics of he UV Cet-type flare stars.
Astrophysics Space Science, {\bf 95}, 235--253 (1983).